# Exploring Proactive Interventions toward Harmful Behavior in Embodied Virtual Spaces

RUCHI PANCHANADIKAR, Clemson University, USA

## 1 INTRODUCTION

Technological advancements have undoubtedly revolutionized various aspects of human life, altering the ways we perceive the world, engage with others, build relationships, and conduct our daily work routines. Among the recent advancements, the proliferation of virtual and mixed reality technologies stands out as a significant leap forward, promising to elevate our experiences and interactions to unprecedented levels. However, alongside the benefits, these emerging technologies also introduce novel avenues for harm and misuse, particularly in virtual and embodied spaces such as Zoom and virtual reality (VR) environments.

The immersive nature of virtual reality environments raises unique challenges regarding psychological and emotional well-being. While VR can offer captivating and immersive experiences, prolonged exposure to virtual environments may lead to phenomena like cybersickness, disorientation, and even psychological distress in susceptible individuals. Additionally, the blurring of boundaries between virtual and real-world interactions in VR raises ethical concerns regarding consent, harassment, and the potential for virtual experiences to influence real-life behavior. Additionally, the increasing integration of artificial intelligence (AI) and machine learning algorithms in virtual spaces introduces risks related to algorithmic bias, discrimination, and manipulation. In VR environments, AI-driven systems may inadvertently perpetuate stereotypes, amplify inequalities, or manipulate user behavior through personalized content recommendations and targeted advertising, posing ethical dilemmas and societal risks.

## 2 MOTIVATION

My previous research has delved into the biases and harms perpetuated by natural language processing (NLP) models towards individuals of different nationalities, as outlined in my publications [1, 2]. While research is actively being conducted to study and mitigate the negative impacts of NLP models and generative AI, the same level of attention has not yet been directed towards newer technologies like virtual reality (VR) and embodied spaces.

I am particularly interested in investigating the harms experienced by female-identifying minority people of color when interacting in these environments. Intersectionality plays a significant role in how biases and harms are propagated and experienced, and studying these dynamics within embodied spaces could provide valuable insights for developing

Author's address: Ruchi Panchanadikar, rapanch@clemson.edu, Clemson University, Clemson, USA.

strategies to mitigate them and create safer spaces for everyone. By focusing on this intersectional perspective, we can work towards establishing guidelines and frameworks that promote inclusivity, equity, and respect in virtual and embodied interactions.

In addition to examining the harms experienced by female-identifying minority people of color in embodied spaces, I am intrigued by the potential of artificial intelligence (AI) to mitigate these harms. Presently, much of the research has concentrated on reactive approaches to harm mitigation, often requiring human intervention after the harm has occurred. However, reactive methods are inherently limited as they address issues only after damage has been done. My interest is on exploring proactive interventions that can anticipate and prevent harmful behavior before it occurs. In this regard, AI holds significant promise. By leveraging AI technologies, we can develop systems capable of detecting and addressing potential harmful behavior in real-time, thereby creating safer and more inclusive spaces for users.